\documentclass[twocolumn,showpacs,preprintnumbers,amsmath,amssymb]{revtex4}

\usepackage{graphicx}
\usepackage{dcolumn}
\usepackage{bm}

\begin{document}

\preprint{JLAB-PHY-02-23}

\title{Single Quark Transition Model Analysis of Electromagnetic
Nucleon Resonance Excitations in the $[70,1^-]$ Supermultiplet.\\}

\author{V.D.~Burkert}
\affiliation{Thomas Jefferson National Accelerator Laboratory, 
Newport News, Virginia 23606, USA\\}

\author{R.~De~Vita, M.~Battaglieri, and M.~Ripani}
\affiliation{Istituto Nazionale di Fisica Nucleare, Sezione di Genova, 16146 Genova, Italy\\}

\author{V.~Mokeev}
\affiliation{Moscow State University, 119899 Moscow, Russia, 
and Christopher Newport University, Newport News, Virginia 23602, 
USA\\}

\date{\today}

\begin{abstract}
We apply the single quark transition model to resonance transition 
amplitudes extracted from photo- and electroproduction data. We 
use experimental data on the $S_{11}(1535)$, and $D_{13}(1520)$ nucleon 
resonances to extract the amplitudes 
for the electromagnetic transition from the nucleon ground state $[56,0^+]$ to the 
 $[70,1^-]$ supermultiplet, and make predictions 
for the transition amplitudes of all other 
states associated with the  $[70,1^-]$. We compare the predictions 
with data and find surprisingly good agreement. 
The comparison is hampered by the poor data quality for many of the 
states especially in the electroproduction sector.     
\end{abstract}

\pacs{ 13.60.le, 13.88.+e, 14.40aq}

\maketitle

\section{Introduction}

Resonance excitation of protons and neutrons is a fundamental 
phenomenon of strong QCD. Descriptions of
resonance excitations have made use of a broad range of constituent 
quark models to describe the excitation of s-channel baryon 
resonances. 
These models are often based on approximate $SU(6)\otimes O(3)$ symmetry 
for the spin-flavor and orbital excitations. This symmetry is broken by 
introducing additional interactions such as the one-gluon 
exchange\cite{isgkar} or Goldstone boson 
exchange between the constituent quarks{\cite{riska}, leading to the 
breaking of mass degeneracies between states belonging to the same 
supermultiplet $[SU(6),L^P]$, where $L^P$ characterizes the orbital 
angular momentum $L$ and parity $P$ of the 3-quark system. The effects of
symmetry breaking on the masses are typically O(150 MeV).    
  
The early successes of the non-relativistic constituent quark 
model in approximately describing many aspects of hadron properties, 
as well as of nucleon structure, led to a broad application 
in electromagnetic interaction. Much effort has gone into describing 
resonance transition amplitudes and form factors \cite{review1}. 
While a reasonable description of many photocoupling amplitudes of low
mass states has been achieved, with the Roper 
resonance $P_{11}(1440)$ being a notable exception, a persistent 
problem in these calculations is the description of the $Q^2$ 
evolution of the transition form factors based on quarks with point like
couplings. This reflects our lack of a full understanding of the concept of 
constituent quarks versus the distance scale probed. Some progress has been 
made in recent years by introducing phenomenological form factors for 
the constituent quarks\cite{simula}, and by going beyond the simple 
harmonic oscillator potential\cite{mauro}. A promising approach 
has been taken recently in employing the field correlator method in 
calculating non-perturbative quark dynamics in baryons\cite{simonov}. In this
approach, constituent quark masses are generated dynamically.  

In the approximation that only a single quark is affected in the 
transition (Single Quark Transition Model, SQTM), simple relationships
can be derived for excitations from the ground state nucleon 
to states assigned to the same $[SU(6),L^P]$ supermultiplet of
the $SU(6)\otimes O(3)$ symmetry group \cite{hey,cottingham,close}. 
Knowledge of only a few amplitudes from states within the same supermultiplet 
allows predictions of amplitudes for all other states within the same
supermultiplet. In this analysis we will assume factorization of 
the spin- and spatial transition matrix elements.  
$SU(6) \otimes O(3)$ symmetry breaking is accomodated by introducing mixing angles 
for states with the same spin, parity, and flavor, but different 
quark spins $S_{3q}=1/2, 3/2$. We use previously determined mixing 
angles from the analysis of hadronic decay properties \cite{mixing1,mixing2}.
Resonance photocoupling amplitudes and their $Q^2$ dependences, including 
their signs, have been 
determined in analyses of pion photo-and electroproduction
experiments taking into account the hadronic couplings as extracted 
from the analyses of hadronic resonance production. We use the signs 
as published which, by convention, are fixed relative to the 
pion Born amplitudes that are included in the analyses aimed at 
extracting resonance photocouplings.  Analysis of these amplitudes
within the SQTM provides information on the consistency of the 
ingredients obtained from hadronic interactions such as the mixing angle. 
The adopted mixing angles can be further tested 
using the $Q^2$-dependence of the amplitudes for the transition to the
$S_{11}(1650)$ and $D_{13}(1700)$ states. Deviations from 
the predicted $Q^2$ dependences may indicate possible violations of
the SQTM assumptions. $SU(6)$ symmetry and the 
single quark transition assumption may be further tested with the 
predictions for other states in the $[70,1^-]$ supermultiplet.      

Experimentally, electromagnetic excitation of nucleon resonances 
(we use nucleon resonances here for both isospin 1/2 and isospin 3/2 
 non-strange baryons), have been studied mostly using 
single pion or eta production. With the new and precise photoproduction 
data which have been collected at MAMI \cite{mami}and GRAAL \cite{graal}, 
and with the new 
photo- and electroproduction data from 
JLab \cite{jlabc,hallc,hallb,jlabb,jlabb1,ripani}, this field is seeing a vast improvement 
in data volume and precision, and much more is expected in various reaction 
channels for the near future. It is therefore timely to revisit 
some of the earlier attempts at coming to a more quantitative 
understanding of electromagnetic resonance excitations.

This paper is organized as follows: In section II we define kinematical 
quantities and the formalism, in section III we briefly review the 
single quark transition model assumptions and summarize the model
predictions. In section IV we review the existing photo- and electroproduction 
data. In section V we present predictions for the transition amplitudes to the 
$[70,1^-]_1$ supermultiplet and compare with the available data. 
Finally we discuss the results in section VI.

\section{Formalism}
  
The inclusive electron scattering cross section is given by
\begin{equation}
{1\over \Gamma_T}{d\sigma \over d\Omega dE^{\prime}} = 
{{1\over 2}(\sigma_T^{1/2} + \sigma_T^{3/2}) + \epsilon_L \sigma_L} \\
\end{equation}

\noindent
where $\Gamma_T$ is the virtual photon flux, $\epsilon_L$ describes the 
degree of longitudinal polarization of the photon, $\sigma_T^{1/2}$ is the total transverse 
absorption cross section with helicity $1/2$ for the photon-nucleon
system, and $\sigma_T^{3/2}$ is the helicity $3/2$ cross section. 
$\sigma_L$ is the total cross section for the absorption of a 
longitudinal photon.  In the following we will focus on the 
transverse part of the cross section as there are insufficient 
data available for a systematic study of the longitudinal couplings of 
nucleon resonances. 
 
In the nucleon resonance region it is convenient to expand the cross section 
for a specific production channel in terms
of partial wave helicity elements. For example, in single pion photoproduction
$\gamma p \rightarrow \pi N$ these are given by:

\begin{eqnarray}
\sigma_T^{1/2} &=& {8\pi q \over k}\sum_{n=0}^{\infty} (n+1)({|A_{n+}|^2 + 
|A_{n+1,-}|^2}) \\\nonumber
\sigma_T^{3/2} &=& {8\pi q \over k}\sum_{n=0}^{\infty} 
{1 \over 4}[n(n+1)(n+2)](|B_{n+}|^2 + |B_{n+1,-}|^2) \\\nonumber
\end{eqnarray}

\noindent
where q and k are, respectively, the pion and photon center-of-mass momenta
calculated at $Q^2=0$, 

\begin{equation}
k = {W^2 - M^2 \over 2W},\\
\end{equation}

\noindent
and 

\begin{equation}
q = \sqrt{[W^2-(m_{\pi}^2-M^2)]-4W^2m_{\pi}^2 \over 2W},\\
\end{equation}

\noindent
and $A_{l\pm}$ and $B_{l\pm}$, with $(l = n,~ n+1)$, are the transverse 
partial wave helicity elements for the pion orbital angular momentum $l$, 
and helicity $1 \over 2$ and $3 \over 2$, respectively. The $l\pm$ 
indicates if the total 
resonance spin is given by $J = l \pm {1\over 2}$, $\pm {1\over 2}$ being 
the nucleon helicity in the initial state. 
These elements are extracted from the experimental data using partial wave analysis techniques. 

For a specific resonance the transverse total photoabsorption cross section 
can be expressed as a function of the transverse photocoupling helicity amplitudes: 

\begin{equation}
\sigma_T^{res} = {2M \over W_R\Gamma}{1\over 2} (A_{1/2}^2 + A_{3/2}^2)\\
\end{equation}

\noindent
where $W_R$ is the resonance mass, and $\Gamma$ the total width.

The $A_{1/2}$, $A_{3/2}$  are related to the partial wave 
helicity elements in the following way:

\begin{eqnarray}
A_{l\pm} &=& \mp f C^I_{\pi^ N} A_{1/2}\\ \nonumber
B_{l\pm} &=& \pm f \sqrt{16 \over (2j-1)(2j+3)} C^I_{\pi N}A_{3/2}\\\nonumber
\end{eqnarray}

\noindent
where

\begin{equation}
f = \sqrt{{1 \over (2j+1)\pi}{k \over q}{M\over W_R}{\Gamma_{\pi}\over 
\Gamma^2}}.\\
\end{equation}

\noindent   
$ C^I_{\pi N}$ are Clebsch-Gordan coefficients describing the 
projection of a resonant state of isospin $I$ into the final state 
$\pi N$ (see table 2). The $A_{l\pm}$, $B_{l\pm}$ can be 
determined directly from experimental data.
Using information from hadronic reactions, the photocoupling helicity 
amplitudes $A_{1/2}$ and $A_{3/2}$ can then be determined. In the following sections
we will use 
these amplitudes to describe the resonance transition to specific states, 
and give the connection to the SQTM amplitudes.\\

\begin{table}
\caption{\label{tab:cpin}C-G coefficients.}
\begin{ruledtabular}
\begin{tabular}{c l l}
{$I$} & {$\pi^+n$} & {$\pi^0p$}\\
\hline
{$\frac{1}{2}$} & {$-\sqrt{\frac{2}{3}}$} & {$+\sqrt{\frac{1}{3}}$}\\ 
{$\frac{3}{2}$} & {$-\sqrt{\frac{1}{3}}$} & {$-\sqrt{\frac{2}{3}}$}\\
\end{tabular}
\end{ruledtabular}
\end{table}

\section{Single Quark Transition Model}\label{sec:sqtm}

Properties of nucleon resonances such as mass, spin-parity, and flavor
 fit well 
into the representation of the $SU(6)\otimes O(3)$ symmetry group, which 
describes the spin-flavor and orbital wave functions of the 3-quark 
system \cite{close}. This symmetry group leads to supermultiplets 
of baryon states with the same orbital angular momentum $L$ of the 
3-quark system, and  degenerate energy levels. Within a supermultiplet 
the quark spins are aligned to form a total quark spin
$S = {1\over 2}$, $3 \over 2$, which combines with the orbital 
angular momentum $L$ to the total angular momentum. 
A large number of explicit dynamical quark models have been developed 
to describe the electromagnetic transition between the nucleon 
ground state and its excited states \cite{warns,capstick,mauro}. 
Measurement of resonance
transitions and the dependence on the distance scales, given by the virtuality 
$Q^2$ of the photon, provides information on the nucleon wave function. 
In order to compute the transition, assumptions on the 3-quark potential and 
the quark-quark interactions have to be made. These are then tested by predicting 
photocoupling helicity amplitudes which can then be confronted with experimental data. 

In this paper we use algebraic relations derived in the literature for resonance 
transitions assuming the transition only affects a single quark in the nucleon.
The parameters in these algebraic equations are then determined 
from experimental analysis. Based on the symmetry properties of the SQTM, 
predictions for a large number of resonance transitions can then be made. 

The fundamentals of the SQTM have been described in 
references \cite{hey,cottingham}, where the symmetry properties 
have been discussed for the transition from the ground state nucleon 
$[56,0^+]$ to the $[70,1^-]$ and the $[56,2^+]$ supermultiplets. 
The $[70,1^-]$  contains states which are prominent in electromagnetic
excitations,  and it is the only supermultiplet for which sufficient
data on resonance couplings of two states are available to extract the 
SQTM amplitudes and test predictions for other states.

The coupling of the electromagnetic current is considered for the 
transverse photon component, and the quarks in the nucleon are assumed to 
interact freely with the photon. 
It has been discussed extensively in the literature 
 \cite{hey,cottingham,close} that in such a model 
the quark transverse current can be written in general as a sum of four terms:

\begin{equation}
J^+= {AL^+ +B\sigma^+L_z  + C\sigma_zL^+  + D\sigma^-L^+L^+}~,\\
\end{equation}

\noindent   
where $\sigma$ is the quark Pauli spin operator, and the terms with 
$A$, $B$, $C$, $D$ in front operate on the quark spatial wave function 
changing the component of orbital angular momentum along the direction of the 
momentum transfer (z- axis). 
The $A$ term corresponds to a quark orbit flip with  $\Delta L_z = +1$, 
term $B$ to a quark spin flip with $\Delta L_z = 0$, 
the $C$ and $D$ terms correspond to simultaneous quark orbit and quark spin flip
with orbital angular momentum flips of $\Delta{L_z} = +1$ 
and $\Delta{L_z} = +2$, respectively. For the transition  from the $[56,0^+]$
to the $[70,1^-]$ supermultiplet with $L = 1$, only $A$, $B$, and $C$ are 
allowed. We limit our discussion to the $[70,1^-]$ supermultiplet as
there is currently insufficient experimental information available 
to extract the SQTM amplitudes for transitions to other supermultiplets.

For the simplest non-relativistic constituent quark model\cite{cko,koniuk},  
only the orbit flip ($A$) and spin flip ($B$) operators are non-zero, 
showing the incompleteness of these descriptions.
The algebraic relations for resonance transitions derived from symmetry 
properties of the SQTM are given in section V. 
Knowledge of 3 amplitudes and of two mixing angles for the transition 
to the $[70,1^-]$ allows predictions for 16 amplitudes of states 
belonging to the same supermultiplet.
If they can be confirmed for some of the 
amplitudes, we have a measure of the degree to which electromagnetic 
transitions of nucleon resonance are dominated by single quark transitions at 
the photon point ($Q^2=0)$ and, using electroproduction data, examine if 
and how this is changing as a function of the distance scale at increasing
photon virtuality.

To present the experimental data we will use the quark electric and magnetic 
multipoles of Cottingham and Dunbar \cite{cottingham}. They provide 
direct physical insight into the resonance transition, and also 
allow simple parameterizations.  
The $A$, $B$, $C$ are simply linear combinations of the quark multipoles. 
For the $[70,1^-]$ multiplet these relations are given by:

\begin{eqnarray}\nonumber
A &=& \;\;\, K \times 2\sqrt{3} e^{11}\\
B &=&     -K \times (\sqrt{6}m^{11} -\sqrt{6}m^{12})\\\nonumber
C &=& \;\;\, K \times (\sqrt{6}m^{11} +\sqrt{6}m^{12}),\\\nonumber
\end{eqnarray}
\noindent

where
\begin{equation}
K = {e\over 2}{1\over \sqrt{M(W^2-M^2)}}.\\
\end{equation}

\noindent
Similar relations have been derived for the transition 
from the ground state $[56,0^+]$ to the $[56,2^+]$ 
supermultiplet. In this case all four SQTM amplitudes contribute 
which, due to the lack of data, currently cannot be determined unambiguously. 
We will, therefore, not discuss 
the transitions to that multiplet here. However, new data from JLab, 
covering a more limited $Q^2$ range in two-pion 
electroproduction\cite{ripani}, may allow determination of 
several states in $[56,2^+]$ in the future.

\subsection{Violation of $SU(6)$ selection rules}

$SU(6)$ symmetry results in  selection rules for transitions to 
some of the states in the $[70,1^-]$ multiplet. For example, 
electromagnetic transitions from proton targets to states in the 
$N^4$ quadruplet with quark spin 
$3\over 2$ are not allowed by the Moorhouse selection rule \cite{moorhouse}. 
$SU(6)$ symmetry is however broken due to 
configuration mixing between various baryon states. Mixing is naturally 
explained as a results of color hyperfine interaction between 
quarks\cite{isgkar} or due to Goldstone boson exchanges\cite{riska}.  
We take these effects into account in the usual way by introducing 
mixing angles for two of the configurations
associated with the $[70,1^-]$ multiplet. Mixing is
present for the $N^2$ and the $N^4$ nucleon states. The  
${1 \over 2}^-$  states are mixed with an angle of $\approx 31^0$, estimated 
from hadronic decay properties \cite{mixing1,mixing2}, 
leading to 
the physical states:

\begin{eqnarray}
|S_{11}(1535)\rangle &=& 0.85\left |N^2, {1 \over 2}^-\right\rangle - 0.53\left |N^4, {1 \over 2}^-\right\rangle\;\;\\\nonumber
|S_{11}(1650)\rangle &=& 0.53\left |N^2, {1 \over 2}^-\right\rangle + 0.85\left |N^4, {1 \over 2}^-\right\rangle,\\\nonumber
\end{eqnarray}

\noindent
where the $|N^2, {1 \over 2}^-\rangle $ and $|N^4, {1 \over 2}^-\rangle $ correspond 
to the nucleon doublet and quadruplet with quark 
spin $1 \over 2$ and $3 \over 2$, respectively.
A much smaller mixing has been observed for the 
$|N^2, {3 \over 2}^-\rangle $ and 
 $|N^4, {3 \over 2}^-\rangle $ states, with a mixing angle of 
$\approx 6^0$, leading to:

\begin{eqnarray}
|D_{13}(1520)\rangle &=& 0.99\left |N^2,{3 \over 2}^-\right\rangle - 0.11\left |N^4, {3 \over 2}^-\right\rangle\;\;\\\nonumber
|D_{13}(1700)\rangle &=& 0.11\left |N^2,{3 \over 2}^-\right\rangle + 0.99\left |N^4, {3 \over 2}^-\right\rangle.\\\nonumber
\end{eqnarray}

As mentioned above, in 
the SQTM $|N^4, {1\over 2}^-> = |N^4, {3\over 2}^-> = 0$
for proton targets. 
However, due to the large mixing angle for the ${1\over 2}^-$ states, 
the SQTM predicts a sizeable  excitation of 
the $S_{11}(1650)$, while the $D_{13}(1700)$ should only 
be weakly excited from proton target. The $D_{15}(1675)$ cannot mix
with any other state, and thus cannot be excited from proton targets with the
SQTM approach.

\section{Experimental helicity amplitudes}  
The test of the SQTM predictions was performed including all photoproduction 
data and all electroproduction data from proton targets presently available.
We did not include electroproduction data from neutron targets as the data 
quality is too poor for a meaningful comparison with our prediction.  
The resonance helicity amplitudes at the photon point were taken from the Particle Data Group~\cite{pdg}. This compilation already combines the outcomes of various analyses, such as the 
ones of Ref.~\cite{awaji, arai, crawford, metcalf}. 
Electroproduction data on the helicity amplitudes are more sparse and available only for the most prominent states. In this analysis we included data from Bonn~\cite{bonn}, DESY~\cite{desy}, NINA~\cite{nina}, and JLab \cite{hallb,hallc}. A compilation of the results obtained at Bonn, DESY, and NINA for $\pi$ and $\eta$ electroproduction can be found in Ref.~\cite{foster}. In addition to the outcomes of the original analysis, we also included the results obtained in the analysis of Ref.~\cite{gerhardt, krosen} and the results of the world data  analysis on $\eta$-electoproduction at the $S_{11}(1535)$ mass presented in Ref.~\cite{hallb}.

\begin{figure}[h]
\vspace{10.cm}
\includegraphics{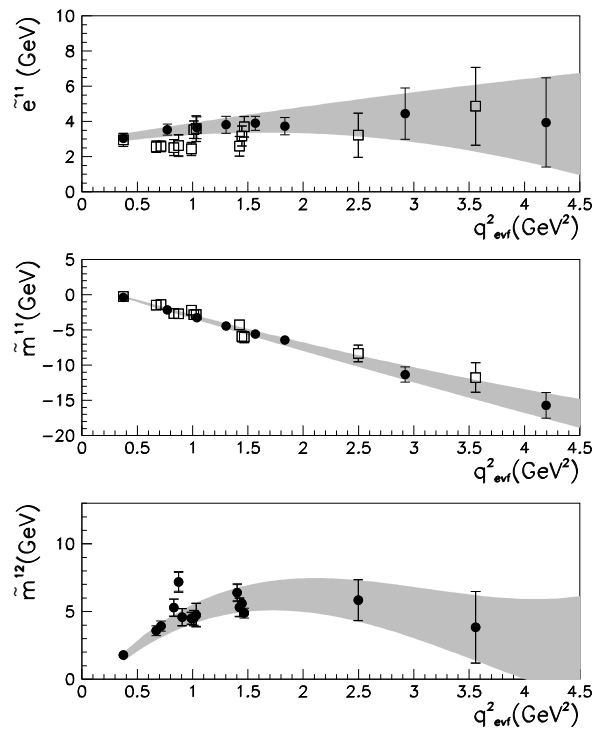}
\caption[]{Extracted quark multipoles as a function of the 
Equal-Velocity-Frame momentum transfer for the $[70,1^-]$ multiplet. 
The shaded band shows the fit results. Their width accounts for the
 uncertainty on the experimental points. In the first two graphs, the 
open squares have been obtained from the Bonn, Nina, and Desy data, 
while the full circles are based on the new JLab measurements. In the 
third plot, all the data are from the Bonn, Nina, and Desy measurements. 
Only the full points were used to derive the quark multipoles 
parameterization.}
\label{fig:qm_70}
\end{figure}
\section{SQTM fit for the $[70,1^-]$ multiplet}
As discussed in Section \ref{sec:sqtm}, the helicity amplitudes of all 
the states that belong to $[70,1^-]$ multiplet can be expressed in terms 
of three SQTM amplitudes at fixed $Q^2$, and the mixing angles obtained from
hadronic resonance decays. 
Therefore three experimentally measured amplitudes are sufficient 
to determine completely the transition from the ground state to 
this multiplet at fixed $Q^2$. For further analysis it is convenient 
to use the quark multipole moments 
$e^{11}$, $m^{11}$, and $m^{12}$ introduced in Ref. \cite{cottingham}. 
They provide direct physical insight and allow simple parametrizations. 
The relations between these quantities and the resonance helicities 
amplitudes can be written as 

\begin{eqnarray}
A_{\frac{1}{2},\frac{3}{2}} = K\left [f_{\frac{1}{2},\frac{3}{2}}^{1} m^{12} + f_{\frac{1}{2},\frac{3}{2}}^{2} e^{11}+ f_{\frac{1}{2},\frac{3}{2}}^{3} m^{11} \right ].
\end{eqnarray}

\noindent
The coefficients $f_{1/2,3/2}^{i}$ for proton and neutron targets are summarized in 
Table \ref{tab:amp_70_proton} and \ref{tab:amp_70_neutron}.
\begin{figure}[h]
\vspace{10.cm}
\includegraphics{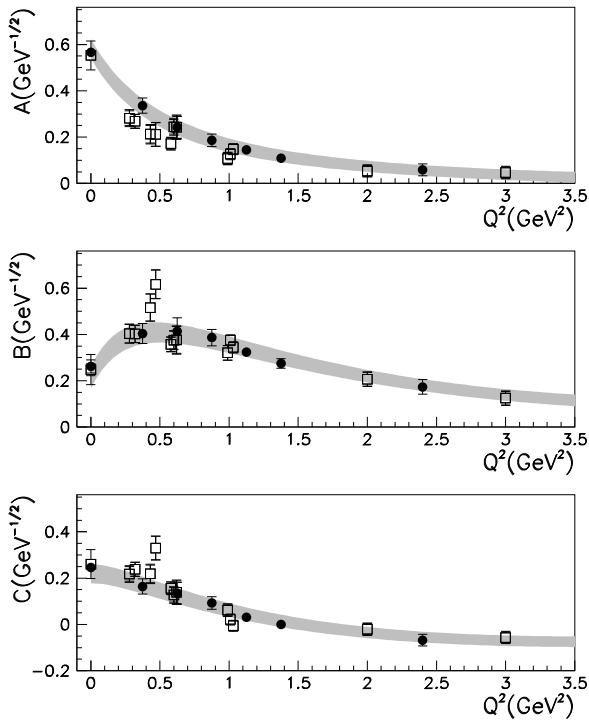}
\caption[]{Single quark transition amplitudes A, B, C for the $[70,1^-]$ multiplet. The shaded band shows the result of the fit of the reduced quark multipoles. The width of the band 
accounts for the uncertainties of the experimental data.}
\label{fig:abc_70}
\end{figure}
To investigate the $Q^2$ dependence of the quark multipoles, the equal velocity frame (EVF) was chosen. In this frame the initial and final hadrons have equal and opposite velocities resulting in minimal relativistic corrections. The four-momentum transfer in the equal velocity frame can be written as

\begin{equation}
{\bf q}^2_{EVF}=\frac{W^2-M^2}{4WM}+Q^2\frac{W^2+M^2}{4WM}.
\end{equation}

\noindent
In order to allow for a simple parametrization of the quark multipoles we 
separate out a common form factor in the $Q^2$ dependence of the resonance 
excitation. We used the usual dipole form: 

\begin{equation}
F({\bf q}^2_{EVF})=(1+{\bf q}^2_{EVF}/0.71)^{-2}
\end{equation}

\noindent
The quark multipoles were then written as

\begin{eqnarray}\nonumber
e^{11} \;\, &=&\tilde e^{11}F({\bf q}_{EVF}^2)\\
m^{11}      &=&\tilde m^{11}F({\bf q}_{EVF}^2)\\\nonumber
m^{12}      &=&\tilde m^{12}F({\bf q}_{EVF}^2),\\\nonumber
\end{eqnarray}

\noindent
where $\tilde e^{11}$, $\tilde m^{11}$, and $\tilde m^{12}$ are called reduced quark multipoles.
The helicity amplitudes for the $S_{11}(1535)$ and $D_{13}(1520)$ resonances, which are the best known states of the $[70,1^-]$ multiplet, were used to derive the quark multipoles from the experimental data. The reduced quark multipoles were then fitted to a smooth curve. The fit results are shown by the shaded band in Figure \ref{fig:qm_70}, where the band width accounts for the uncertainty of the measured amplitudes. The central values of the band can be written analytically as follows ({\bf q}$_{EVF}^2$ $<$ 4 GeV$^2$):

\begin{eqnarray}\nonumber
\tilde e^{11} \;\, &=& \;\;\, 2.67 \;+\; 1.10 \;{\bf q}_{EVF}^2 \;-\; 0.21 \;{\bf q}_{EVF}^4\\
\tilde m^{11}      &=& \;\;\, 1.32 \;-\; 4.51 \;{\bf q}_{EVF}^2 \;-\; 0.10 \;{\bf q}_{EVF}^4\\\nonumber
\tilde m^{12}      &=&       -1.34 \;+\; 9.20 \;{\bf q}_{EVF}^2 \;-\; 3.39 \;{\bf q}_{EVF}^4\\\nonumber
                 &\phantom{=}& \;\;\;\;\;\;\;\;\;\;\; + 0.34 \;{\bf q}_{EVF}^6. \\\nonumber
\end{eqnarray}

\noindent
The ${\bf q}_{EVF}^6$ term was added as the $\tilde m^{12}$ multipole shows a more complicated 
dependence on ${\bf q}_{EVF}$  than the 
other multipoles.  The quark multipole moments were then used to 
evaluate the SQTM prediction
 for all the states of the  $[70,1^-]$ multiplet. The results are shown in 
Figure \ref{fig:sqtm_70_p} and \ref{fig:sqtm_70_n}, where the SQTM predictions represented by the shaded 
bands are compared with the data.  
We do not show predictions for the $D_{15}(1675)$ on the proton as the state cannot
mix and its amplitudes are predicted to be zero. \\

\begin{figure*}
\vspace{12.cm}
\includegraphics{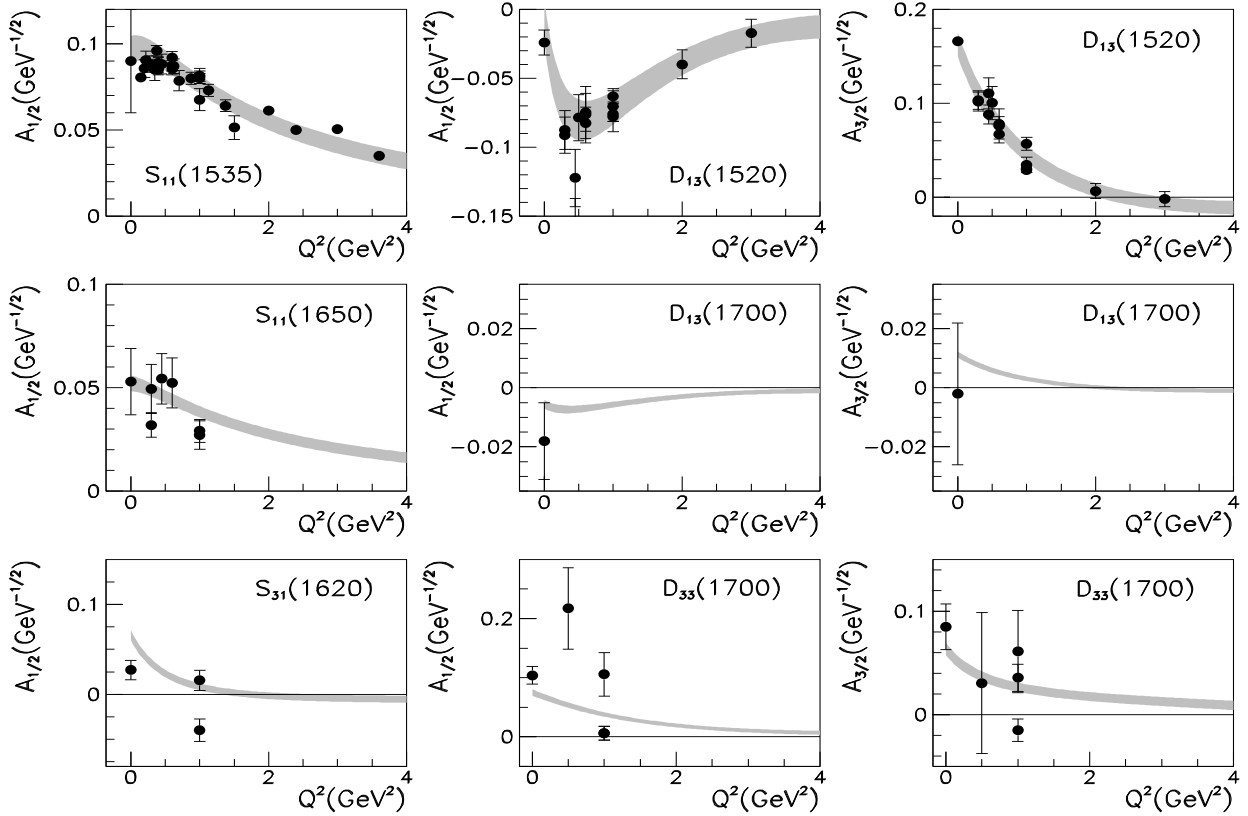}
\caption[]{Single quark model prediction for the $[70,1^-]$ multiplet on the proton. 
The SQTM predictions are shown by the 
shaded band in comparison with the experimental data. At $Q^2=0$ the full 
circle is the Particle Data Group estimate. For $Q^2>0$, measurements from 
JLab, Bonn, DESY, and NINA in $\eta$ and $\pi$ electroproduction are shown. 
For the $S_{11}(1535)$, the results of an analysis of the world data 
in $\eta$-electroproduction presented in Ref. \cite{hallb} are also included.}
\label{fig:sqtm_70_p}
\end{figure*}

\begin{figure*}
\vspace{12.cm}
\includegraphics{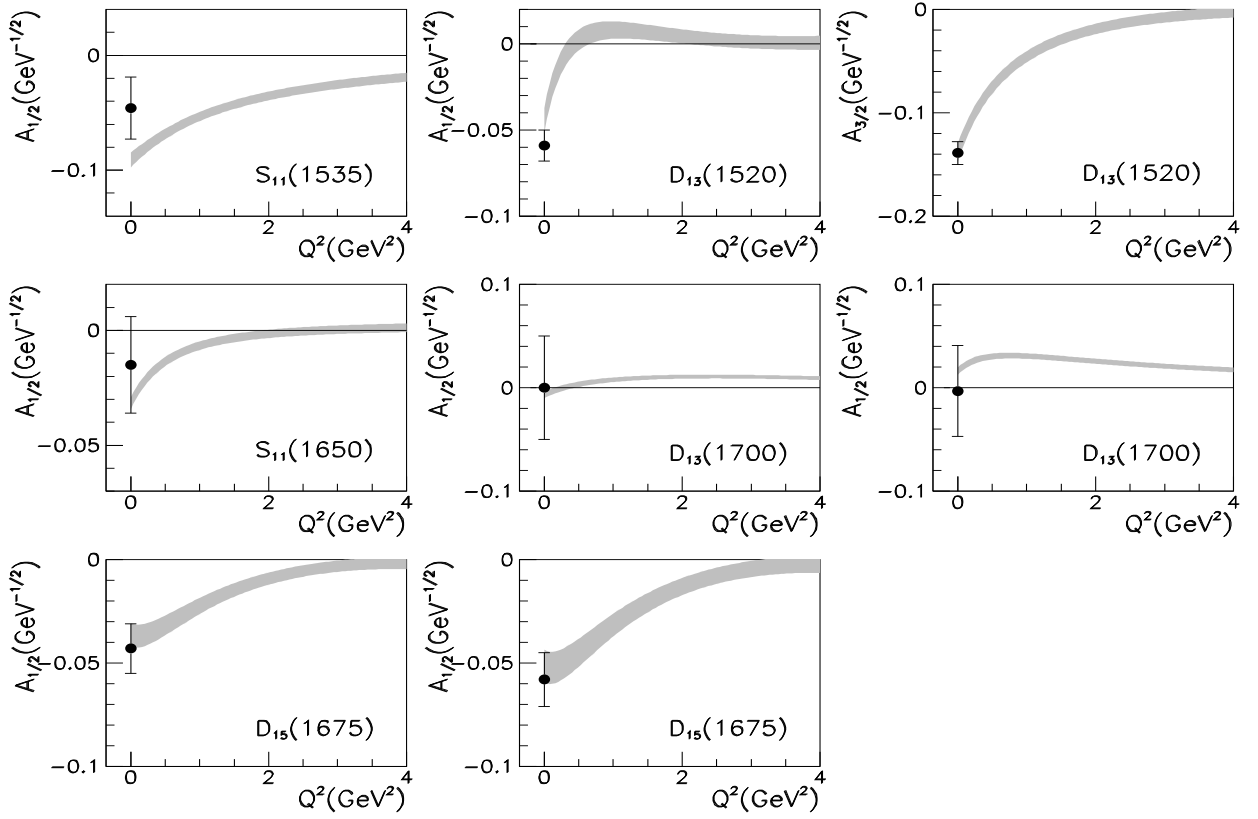}
\caption[]{Single quark model prediction for the $[70,1^-]$ multiplet on the 
neutron. 
The SQTM predictions are shown by the 
shaded band in comparison with the experimental data. At $Q^2=0$ the full 
circle is the Particle Data Group estimate.}
\label{fig:sqtm_70_n}
\end{figure*}

\begin{table*}
\caption{\label{tab:amp_70_proton} Helicity amplitudes of the $[70,1^-]$ multiplet on a proton target as a function of the quark multipoles $e^{11}$, $m^{11}$, and  $m^{12}$.}
\begin{ruledtabular}
\begin{tabular}{c c r r r r r}
{State} & {Amplitude} & {$f^{1}$} & {$f^{2}$} & {$f^{3}$} \\
\hline
\hline
{S$_{11}$(1535)} & {$A_{1/2}$} & {$ \sqrt{\frac{1}{3}}\cos 31^{\circ}$}  & {$-\sqrt{\frac{2}{3}} \cos 31^{\circ}$}  & {}  \\
{D$_{13}$(1520)} & {$A_{1/2}$} & {$ \sqrt{\frac{1}{6}}\cos  6^{\circ}$}  & {$ \sqrt{\frac{1}{12}}\cos  6^{\circ}$}  & {$-\sqrt{\frac{3}{4}}\cos  6^{\circ}$}  \\ 
{}               & {$A_{3/2}$} & {$ \sqrt{\frac{1}{2}}\cos  6^{\circ}$}  & {$ \frac{1}{2}        \cos  6^{\circ}$}  & {$ \frac{1}{2}       \cos  6^{\circ}$}  \\  
{S$_{11}$(1650)} & {$A_{1/2}$} & {$ \sqrt{\frac{1}{3}}\sin 31^{\circ}$}  & {$-\sqrt{\frac{2}{3}} \sin 31^{\circ}$}  & {}  \\  
{S$_{31}$(1620)} & {$A_{1/2}$} & {$ \sqrt{\frac{1}{3}}$}                 & {$\sqrt{\frac{2}{27}}$}                  & {}  \\  
{D$_{15}$(1675)} & {$A_{1/2}$} & {}  & {}  & {}  \\  
{}               & {$A_{3/2}$} & {}  & {}  & {}  \\  
{D$_{13}$(1700)} & {$A_{1/2}$} & {$ \sqrt{\frac{1}{6}}\sin  6^{\circ}$}  & {$ \sqrt{\frac{1}{12}}\sin  6^{\circ}$}  & {$-\sqrt{\frac{3}{4}}\sin  6^{\circ}$}  \\
{}               & {$A_{3/2}$} & {$ \sqrt{\frac{1}{2}}\sin  6^{\circ}$}  & {$ \frac{1}{2}	 \sin  6^{\circ}$}  & {$ \frac{1}{2}	   \sin  6^{\circ}$}  \\
{D$_{33}$(1700)} & {$A_{1/2}$} & {$ \sqrt{\frac{1}{6}}$}                 & {$-\frac{1}{2\sqrt{27}}$}                & {$\sqrt{\frac{1}{12}}$}  \\  
{}               & {$A_{3/2}$} & {$ \sqrt{\frac{1}{2}}$}                 & {$-\frac{1}{6}$}                         & {$-\frac{1}{6}$}  \\  
\end{tabular}
\end{ruledtabular}
\end{table*}

\begin{table*}
\caption{\label{tab:amp_70_neutron} Helicity amplitudes of the $[70,1^-]$ multiplet on a neutron target as a function of the quark multipoles $e^{11}$, $m^{11}$, and  $m^{12}$.}
\begin{ruledtabular}
\begin{tabular}{c c r r r r r}
{State} & {Amplitude} & {$f^{1}$} & {$f^{2}$} & {$f^{3}$} \\
\hline
\hline
{S$_{11}$(1535)} & {$A_{1/2}$} & {$-\sqrt{\frac{1}{3}}\cos 31^{\circ}$}  & {$\sqrt{\frac{2}{27}}\cos 31^{\circ}-\sqrt{\frac{2}{27}}\sin 31^{\circ}$}              & {}  \\  
{D$_{13}$(1520)} & {$A_{1/2}$} & {$-\sqrt{\frac{1}{6}}\cos  6^{\circ}$}  & {$-\frac{1}{2\sqrt{27}}\cos  6^{\circ}+\frac{1}{2}\sqrt{\frac{10}{27}}\sin 6^{\circ}$} & {$\sqrt{\frac{1}{12}}\cos 6^{\circ}+\sqrt{\frac{1}{30}}           \sin 6^{\circ}$}  \\
{}	         & {$A_{3/2}$} & {$-\sqrt{\frac{1}{2}}\cos  6^{\circ}$}  & {$-\frac{1}{6}         \cos  6^{\circ}+\frac{\sqrt{10}}{6}		 \sin 6^{\circ}$} & {$-\frac{1}{6}	 \cos 6^{\circ}-\frac{1}{3}\frac{1}{\sqrt{10}}\sin 6^{\circ}$}  \\
{S$_{11}$(1650)} & {$A_{1/2}$} & {$-\sqrt{\frac{1}{3}}\sin 31^{\circ}$}  & {$\sqrt{\frac{2}{27}}\sin 31^{\circ}-\sqrt{\frac{2}{27}}\cos 31^{\circ}$}              & {}  \\  
{S$_{31}$(1620)} & {$A_{1/2}$} & {$\sqrt{\frac{1}{3}}$}                  & {$\sqrt{\frac{2}{27}}$}                                                                & {}  \\  
{D$_{15}$(1675)} & {$A_{1/2}$} & {}                                      & {$-\frac{1}{2\sqrt{27}}$}				                                  & {$-\sqrt{\frac{2}{15}}$}  \\
{}		 & {$A_{3/2}$} & {}                                      & {$-\frac{1}{6}	  $}				                                  & {$-\sqrt{\frac{4}{15}}$}  \\
{D$_{13}$(1700)} & {$A_{1/2}$} & {$-\sqrt{\frac{1}{6}}\sin  6^{\circ}$}  & {$-\frac{1}{2\sqrt{27}}\sin  6^{\circ}-\frac{1}{2}\sqrt{\frac{10}{27}}\cos 6^{\circ}$} & {$\sqrt{\frac{1}{12}}\sin 6^{\circ}-\sqrt{\frac{1}{30}}           \cos 6^{\circ}$}  \\
{}		 & {$A_{3/2}$} & {$-\sqrt{\frac{1}{2}}\sin  6^{\circ}$}  & {$-\frac{1}{6}	  \sin  6^{\circ}-\frac{\sqrt{10}}{6}	         \cos 6^{\circ}$} & {$-\frac{1}{6}	 \sin 6^{\circ}+\frac{1}{3}\frac{1}{\sqrt{10}}\cos 6^{\circ}$}  \\
{D$_{33}$(1700)} & {$A_{1/2}$} & {$\sqrt{\frac{1}{6}}$}                  & {$-\frac{1}{2\sqrt{27}}$}                                                              & {$\sqrt{\frac{1}{12}}$}  \\  
{}		 & {$A_{3/2}$} & {$\sqrt{\frac{1}{2}}$}                  & {$-\frac{1}{6}$}                                                                       & {$-\frac{1}{6}$}  \\ 
\end{tabular}
\end{ruledtabular}
\end{table*}

\section{Discussion}

Comparison of the SQTM predictions with the data shows globally
good agreement with the sparse data indicating that the
model accounts for the main features of the excitations to the $[70,1^-]$ for
$Q^2 \leq 1$ GeV$^2$.} We find that for states for which 
the signs of the 
amplitudes are known they are correctly predicted by the SQTM analysis. 
There is also quantitative agreement for most amplitudes at the photon 
point, and for 
amplitudes where good data on the $Q^2$ evolution are available, e.g. for 
the $S_{11}(1650)$ on the proton,  
excellent agreement is seen as well. The SQTM predictions, using a $31^o$ 
mixing angle agree both in magnitude at the photon point, as well as with
the $Q^2$ dependence. We take this as a confirmation of the adopted 
mixing angle, and as an indication that the SQTM works at a reasonable 
level for this state. Good agreement is also seen at the photon point 
for most other states. The poor quality of the electroproduction data for 
all other states does not allow drawing more definit conclusions. 

In the case of the neutron states, both of the $D_{13}(1700)$ amplitudes, 
as well as the $D_{13}(1520)$ amplitudes, agree very well with the 
predicted ones at the photon point, while there are no electroproduction 
data available. 
Similarly, both $D_{15}(1675)$ amplitudes are in good agreement with 
the photoproduction data. This may be interpreted as a more direct 
confirmation of the SQTM assumptions as the latter state is not 
affected by mixing.

In the case of the $D_{33}(1700)$, 
the very large value of $A_{1/2}$ at $Q^2=0.5$ GeV$^2$ is likely unphysical, 
as it would produce 
a prominent enhancement in the inclusive cross section, which is not 
seen in the data. The $Q^2 = 1$ GeV$^2$ points are in disagreement with each other, 
while their average agrees with our prediction. A similar discrepancy between two  
data sets is seen for the 
$Q^2 = 1$ GeV$^2$ points of the $S_{31}(1620)$. Our prediction agrees with one of them. 
Given such systematic uncertainties in the 
electroproduction data we conclude that the SQTM predictions 
for the electromagnetic transition from the nucleon ground state to the $[70,1^-]$ 
supermultiplet compare favorably with the available data. 
Obviously,  much improved data are needed for more stringent tests of the model 
assumptions. 
It will be very interesting to see if and where the SQTM predictions break down. 
Such a breakdown
could be due to non-quark contributions at lower $Q^2$, for example pion cloud effects. 
Such effects are currently being studied\cite{morel}. It may also indicate sizeable 
multi-quark transitions.  

Additional experimental information on at least one state 
in the $[56,2^+]$ supermultiplet, 
e.g. the $P_{13}(1720)$, is needed to uniquely extract the SQTM amplitudes for 
that supermultiplet. The main reason for the lack of data for states in the 
 $[70,1^-]$ and $[56,2^+]$ supermultiplets is that many states couple only weakly to 
the $N\pi$ channel, the main source of information on resonance excitations. Most states have 
rather strong couplings to the $N\pi\pi$ channels. These channels are currently being 
studied in several experiments at JLab, GRAAL, and ELSA.  Recent measurements of 2-pion 
electroproduction at JLab may give access to several other states 
assigned to the $[56,2^+]$ supermultiplet. 
This will allow more stringent tests of 
the SQTM predictions than are currently possible. 
No photo- or electroproduction data exist for any 
of the higher supermultiplets. \\

\end{document}